\documentclass[sigconf,review=false,screen=false,authorversion=true,anonymous=false]{acmart}

\usepackage{graphicx}  %

\usepackage{multirow}
\usepackage{algorithm,algpseudocode}
\usepackage{subcaption}
\usepackage{amssymb}
\usepackage{enumitem}
\usepackage{float}
\newcommand{\tabincell}[2]{\begin{tabular}{@{}#1@{}}#2\end{tabular}}

\AtBeginDocument{%
  \providecommand\BibTeX{{%
    \normalfont B\kern-0.5em{\scshape i\kern-0.25em b}\kern-0.8em\TeX}}}

\copyrightyear{2020} 
\acmYear{2020} 
\setcopyright{acmlicensed}\acmConference[MM '20]{Proceedings of the 28th ACM International Conference on Multimedia}{October 12--16, 2020}{Seattle, WA, USA}
\acmBooktitle{Proceedings of the 28th ACM International Conference on Multimedia (MM '20), October 12--16, 2020, Seattle, WA, USA}
\acmPrice{15.00}
\acmDOI{10.1145/3394171.3413721}
\acmISBN{978-1-4503-7988-5/20/10}

\acmSubmissionID{1722}

\begin{document}
\fancyhead{}
\title{PopMAG: Pop Music Accompaniment Generation}

\author{Yi Ren$^{1*}$, Jinzheng He$^{1*}$, Xu Tan$^{2}$, Tao Qin$^{2}$, Zhou Zhao$^{1\dagger}$, Tie-Yan Liu$^{2}$}
\affiliation{$^{1}$Zhejiang University, $^{2}$Microsoft Research Asia}
\email{{rayeren,3170106086}@zju.edu.cn,{xuta,taoqin}@microsoft.com,zhaozhou@zju.edu.cn,tyliu@microsoft.com}

\begin{abstract}
In pop music, accompaniments are usually played by multiple instruments (tracks) such as drum, bass, string and guitar, and can make a song more expressive and contagious by arranging together with its melody. Previous works usually generate multiple tracks separately and the music notes from different tracks not explicitly depend on each other, which hurts the harmony modeling. To improve harmony, in this paper\footnote{$^*$ Equal contribution. $\dagger$ Corresponding author.}, we propose a novel MUlti-track MIDI representation (MuMIDI), which enables simultaneous multi-track generation in a single sequence and explicitly models the dependency of the notes from different tracks. While this greatly improves harmony, unfortunately, it enlarges the sequence length and brings the new challenge of long-term music modeling. We further introduce two new techniques to address this challenge: 1) We model multiple note attributes (e.g., pitch, duration, velocity) of a musical note in one step instead of multiple steps, which can shorten the length of a MuMIDI sequence. 2) We introduce extra long-context as memory to capture long-term dependency in music. We call our system for pop music accompaniment generation as PopMAG. We evaluate PopMAG on multiple datasets (LMD, FreeMidi and CPMD, a private dataset of Chinese pop songs) with both subjective and objective metrics. The results demonstrate the effectiveness of PopMAG for multi-track harmony modeling and long-term context modeling. Specifically, PopMAG wins 42\%/38\%/40\% votes when comparing with ground truth musical pieces on LMD, FreeMidi and CPMD datasets respectively and largely outperforms other state-of-the-art music accompaniment generation models and multi-track MIDI representations in terms of subjective and objective metrics.

\end{abstract}


\maketitle

\section{Introduction}
Music generation~\cite{dong2018musegan,yang2017midinet,huang2018music,huang2020pop,nierhaus2009algorithmic,johnson2017generating} has attracted a lot of attention in both research, industrial and art community in recent years. Similar to natural language, a music sequence is usually represented as a series of symbolic tokens (e.g., MIDI) and modeled with deep learning techniques, including CNN~\cite{yang2017midinet}, RNN~\cite{simon2017performance}, Transformer~\cite{huang2018music,huang2020pop,child2019generating}, VAE~\cite{roberts2018hierarchical} and GAN~\cite{dong2018musegan}. In pop music, the generation of a song usually consists of 1) chord and melody generation; and 2) accompaniment generation based on chord and melody. In this paper, we focus on accompaniment generation. Considering music accompaniments usually introduce multiple instruments/tracks (e.g., guitar, bass, drum, piano and string in pop music) in arrangement for better expressiveness, we also call it as multi-track music generation. 

A key problem for multi-track generation is how to ensure harmony among the musical notes in multiple tracks. Previous works~\citep{dong2017musegan,dong2018musegan,dong2018convolutional,zhu2018xiaoice,liang2019midi,donahue2019lakhnes} have tried to keep harmony among all generated music tracks. MuseGAN~\cite{dong2017musegan,dong2018musegan,dong2018convolutional} generates music as an image (converting MIDI (a digital score format) into pianoroll) with generative adversarial networks (GANs), and uses an inter-track latent vector to make the generated music coherent. MIDI-Sandwich2~\cite{liang2019midi}, which also uses pianoroll-based MIDI representations, applies a multi-modal simultaneous generation method to combine individual RNN to collaboratively generate harmonious multi-track music. However, pianoroll-based generation is unstable to train and suffers from data sparseness, which makes the quality of generated music worse than the level of human musicians. XiaoIce Band~\cite{zhu2018xiaoice} introduces cooperate cells between each generation track to ensure harmony. However, the dependency among the musical notes in different tracks in the same generation step is missing in XiaoIce Band. %
LakhNES~\cite{donahue2019lakhnes} uses different tokens to represent the note of different instruments, which makes it hard to model the relationship of the same pitch among instruments and thus affects the harmony.

\begin{figure*}[!h]
    \centering
    \vspace{-0.0cm}
	\includegraphics[width=0.95\textwidth, trim={0cm 9.7cm 0cm 0cm}, clip=true]{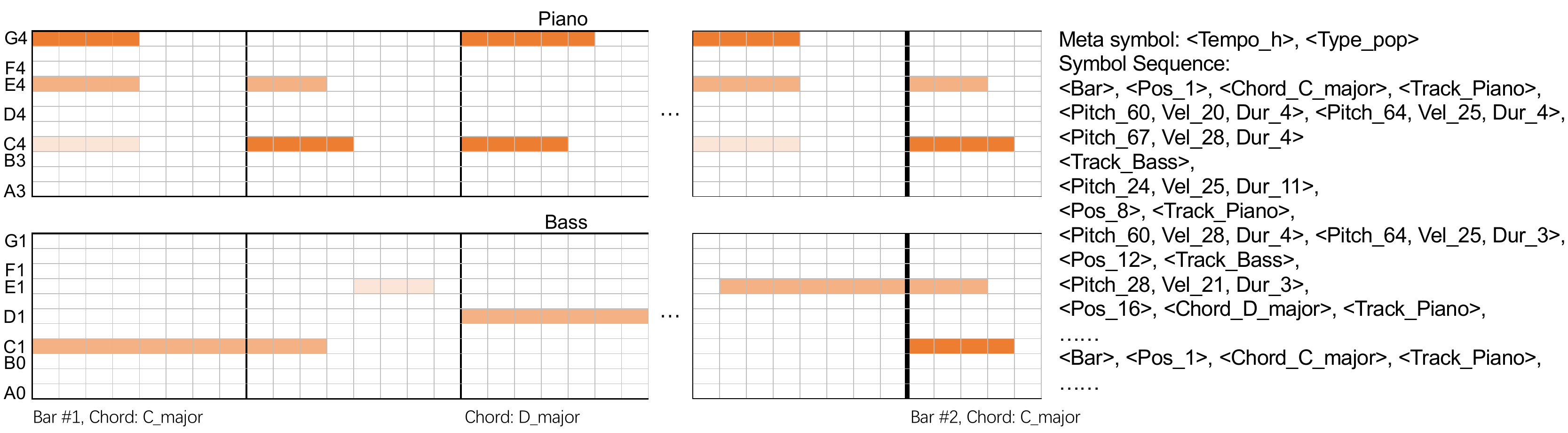}
	\vspace{-0.3cm}
	\caption{An example of a MuMIDI token sequences converted from a segment of music score. The musical piece in left subfigure is represented as a sequence of symbols as shown in right subfigure.}
	\label{fig:mumidi_example}
	\vspace{-0.5cm}
\end{figure*}

In this paper, to ensure harmony in music accompaniment generation, we propose a novel MUlti-track MIDI representation (MuMIDI) that encodes multi-track MIDI events into a sequence of tokens. Instead of generating multiple tracks separately in previous works, MuMIDI enables multi-track generation in a single sequence. In this way, the dependency among the musical notes in different tracks can be better captured and more information can be leveraged to improve harmony: the generation of a music note in one track at $t$-th step explicitly depends on (1) the notes generated at $<t$ steps in the same track, (2) the notes generated at $<t$ steps in all other tracks, and (3) the notes at $t$-th step for tracks that have been generated. 

Music relies heavily on repetition and long-term structure to make a musical piece coherent and understandable. Unfortunately, MuMIDI encodes multi-track MIDI events into a single sequence and increases the length of the music sequence, consequently increasing the difficulty of long-term music modeling. To address this challenge, we adopt a sequence-to-sequence model and enhance it from two aspects to better model long-term sequences: 1) We shorten the sequence length by modeling multiple attributes (e.g., pitch, duration, velocity) that belong to a music note in one sequence step instead of multiple steps. 2) We adopt Transformer-XL~\cite{dai2019transformer}  as the backbone of the encoder and decoder of our sequence-to-sequence model to capture the long-term dependencies. The encoder takes the tokens in conditional tracks (e.g., chord and melody) as input and the decoder predicts the tokens in target tracks in an autoregressive manner.
In this way, our model can fuse the information from both input tracks (e.g., chord and melody) and previously generated notes in target tracks to generate harmonious music and memorize the long-term music structure as well.

To summarize, our designed system for Pop Music Accompaniment Generation (PopMAG for short) contains two key technical parts: MuMIDI representations and enhanced sequence-to-sequence model. We test PopMAG on three pop music datasets and evaluate the quality of generated music with both objective metrics and subjective testing. The results show that PopMAG wins 42\%/38\%/40\% votes when comparing with ground truth musical pieces on LMD~\cite{raffel2016learning}, FreeMidi\footnote{https://freemidi.org/genre-pop} and CPMD~\footnote{Our internal chinese pop MIDI dataset.} datasets respectively and largely outperforms state-of-the-art music accompaniment generation systems. In particular, we observe that 1) compared with previous representations such as MIDI and REMI~\cite{huang2020pop}, MuMIDI shows great advantages in modeling harmony; and 2) our proposed sequence-to-sequence model can better capture long-term dependencies. Our generated music samples can be found in \url{https://music-popmag.github.io/popmag/}.

\section{Multi-track MIDI Representation}

Music accompaniments usually contain multiple tracks performed at the same time, where the harmony between tracks is important for music quality. How to ensure harmony among multiple tracks is important to generate high-quality music accompaniments. Some previous works~\cite{zhu2018xiaoice,dong2018musegan,dong2018convolutional} generate each track in separate decoder and ensure their consistency by imposing adversarial loss or implicit latent vector. However, they cannot model the dependency among the music notes in different tracks explicitly. Simply combining multiple tracks together like pianoroll~\cite{dong2018musegan,dong2018convolutional} cannot ensure harmony and suffers from data sparseness. In this section, we design a novel Multi-track MIDI representation (MuMIDI) to encode multi-track music notes into a single compact sequence of tokens. In this way, the harmony between different tracks can be modeled inherently using MuMIDI.

In MuMIDI, to encode a multi-track musical piece with complex data structure into a single sequence of symbols, we introduce some symbols including bar, position, track, note, chord and meta symbols. Among them, bar and position symbols together represent the beginning of a new position in the musical piece, followed by a track symbol denoting the beginning of a new track or a chord symbol denoting the chord that subsequent notes should follow, and finally note symbols are added. Figure \ref{fig:mumidi_example} shows an example of MuMIDI. The left subfigure is a 2-track musical piece, and the corresponding symbol sequence in MuMIDI is shown in the right subfigure. This musical piece contains two tracks: piano track and bass track, which contain 10 notes and 5 notes, respectively. The sequence starts with a bar symbol (<\textit{Bar}>), followed by a position symbol (<\textit{Pos\_1}>), a chord symbol (<\textit{Chord\_C\_major}>) and a piano track symbol (<\textit{Track\_Piano}>). Then 3 note symbols are added, each of them contains 3 attributes (\textit{Pitch}, \textit{Velocity} and \textit{Duration}). When the track changes, a new track symbol (<\textit{Track\_Bass}>) needs to be added indicating the track to which subsequent notes belong. We introduce each kind of symbol in MuMIDI in the following subsections.

\subsection{Bar and Position}
\label{sec:bar_pos}
Inspired by REMI~\cite{huang2020pop}, we use bar and position symbols to indicate the beginning of bar and different positions in a bar.

\paragraph{Bar} We use a <$Bar$> symbol to indicate the beginning of each bar. 
All symbols in the bar (including the position symbols, track symbols, note symbols and chord symbols) will be added after the <$Bar$> symbol at the beginning of the bar. When a new bar begins, another <$Bar$> symbol is added.

\paragraph{Position} We divide a bar into 32 timesteps evenly and quantize the onset time of each note to the nearest timestep. We use <\textit{Pos\_1}>, <\textit{Pos\_2}>, ..., <\textit{Pos\_32}> to represent the beginning of 32 timesteps. Different from REMI~\cite{huang2020pop} which appends one position symbol before each note or chord symbols, we only use one position symbol to represent the beginning of each timestep: all other symbols (including track symbols, note symbols and chord symbols) starting at this timestep are appended after this position symbol. This modification can shorten the symbol sequence and help maintain long-term memory better in multi-track scenario.

\subsection{Track}
After each position symbol, a track symbol is appended to indicate the track to which subsequent notes belong. In this paper, we use 6 types of track symbols,  <\textit{Track\_Melody}>, <\textit{Track\_Drum}>, <\textit{Track\_Piano}>, <\textit{Track\_String}>, <\textit{Track\_Guitar}>, <\textit{Track\_Bass}>, to represent melody, drum, piano, string, guitar, and bass track respectively. We append all note symbols belong to this track after the track symbol. When the track changes, another track symbol will be added. In this way, we can put all tracks of notes in one sequence to keep harmonious among all tracks.

\subsection{Note}
A note symbol has some attributes including pitch, velocity and duration. ``MIDI-like'' representation~\cite{huang2018music} use MIDI events, such as \textit{Set Velocity}, \textit{Note On}, \textit{Time Shift} and \textit{Note Off} to describe a note, while REMI~\cite{huang2020pop} replaces the \textit{Note Shift} with \textit{Note Duration} to make the duration explicit and facilitate modeling the rhythm of notes. However, those commonly-adopted representation uses three or more tokens to represent a note, making the token sequence extremely long. In our representation, all attributes of a note will be represented in one symbol. We list all note attributes in Table \ref{tab:note_attr}. \textit{Pitch} attribute indicates note pitches from 1 (C-1) to 128 (G9) for all tracks except drum. \textit{Drum Type} attribute indicates the drum type of notes for drum track following the percussion instrument mappings defined in General MIDI protocol\footnote{\url{https://en.wikipedia.org/wiki/General\_MIDI}}. For \textit{Velocity}, we quantize the velocity into 32 levels, corresponding to \textit{Velocity\_1} to \textit{Velocity\_32}. \textit{Duration} attribute indicates the duration of note from 1 timestep to 32 timesteps.

\begin{table}[!h]
\vspace{-0.4cm}
\caption{Note attributes in MuMIDI.}
\vspace{-0.4cm}
\small
	\centering
	\begin{tabular}{ l | c | c  }
	\toprule
		Attribute Name &  Representation  & \# Size  \\
		\midrule
		\textit{Pitch} & \textit{Pitch\_1}, \textit{Pitch\_2},..., \textit{Pitch\_128} & 128  \\ \hline
		\textit{Drum Type} & \textit{Drum\_1}, \textit{Drum\_2},..., \textit{Drum\_128} & 128  \\ \hline
		\textit{Velocity} & \textit{Vel\_1}, \textit{Vel\_2},..., \textit{Vel\_32} & 32  \\ \hline
		\textit{Duration} & \textit{Dur\_1}, \textit{Dur\_2},..., \textit{Dur\_32} & 32  \\ 
		\bottomrule
	\end{tabular}
\label{tab:note_attr}
\vspace{-0.4cm}
\end{table}

\subsection{Chord}

Chord symbols are a set of auxiliary musical symbols to represent the chord progression which guides the pitch range of notes and emotion and is very important in pop music composition. Chord progression changes over time but does not contain any real note. A chord usually consists of a root note and a chord quality~\cite{mcfee2017structured}. In MuMIDI, we consider 12 chord roots ($\textit{C}$,$\textit{C\#}$,$\textit{D}$,$\textit{D\#}$,$\textit{E}$,$\textit{F}$,$\textit{F\#}$,$\textit{G}$,$\textit{G\#}$,$\textit{A}$,$\textit{A\#}$,$\textit{B}$) and 7 chord qualities ($\textit{major}$, $\textit{minor}$, $\textit{diminished}$, $\textit{augmented}$, $\textit{major7}$, $\textit{minor7}$, $\textit{half\_diminished}$), resulting in totally 84 possible chord symbols. A chord symbol is usually appended to a position symbol before a track symbol, indicating the chord which subsequent notes in all tracks should follow. We assume that each chord remains unchanged for half a bar, and therefore the chord symbol is only appended after <\textit{Pos\_1}> and <\textit{Pos\_16}>.

\subsection{Meta Symbol}
Meta symbols encode the meta data of the whole musical piece, such as tempo, tonality, style and emotion, which is usually unchanged throughout the whole musical piece. In this paper, as a demonstration, we only use tempo as the meta symbol: we simply classify the tempo into three categories: low (lower than 90), middle (90 to 160) and high (higher than 160) and use three meta symbol to represent them. Although other meta symbols will not be discussed in detail in this paper, we can easily implement style/emotion-controllable music accompaniment generation in our framework with meta symbols.

\section{Multi-Track Modeling}

MuMIDI encodes multi-track MIDI events into a single sequence, which could be very long and cause difficulty for long-term structure modeling in music. Therefore, we adopt a sequence-to-sequence model and enhance it from two aspects to better model long-term sequences: 1) We shorten the sequence length by modeling multiple note attributes (e.g., pitch, duration, velocity) that belong to a musical note in one sequence step instead of multiple steps. 2) We adopt extra long context as used in Transformer-XL~\cite{dai2019transformer} in the encoder and decoder of our sequence-to-sequence model to capture long-term dependencies. In the next subsections, we first introduce the above two enhancements and then describe the overall implementation of the enhanced sequence-to-sequence model.

\begin{figure}
    \centering
    \includegraphics[width=0.41\textwidth, trim={0cm 9.0cm 0.0cm 0cm}, clip=true]{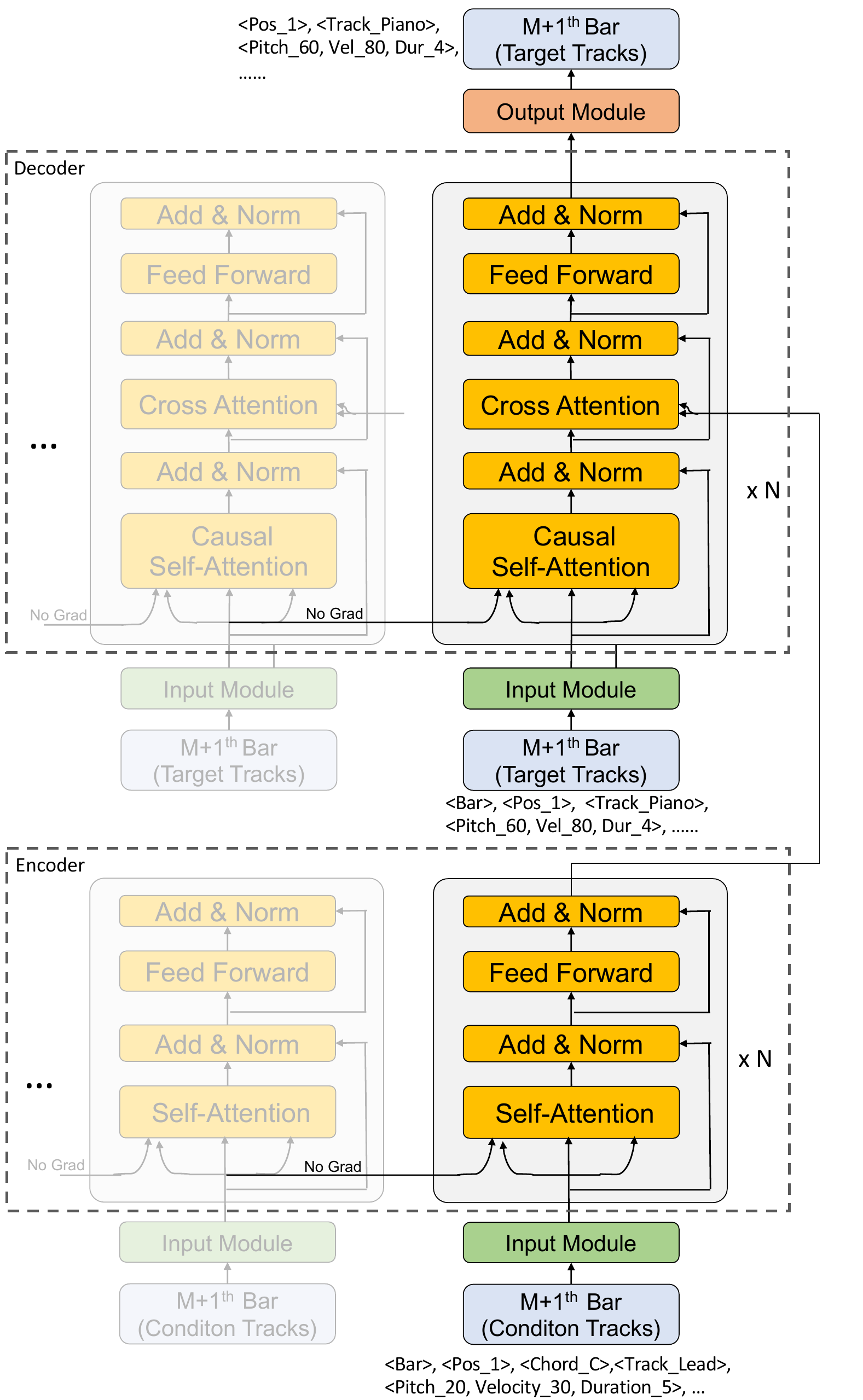}
    \caption{The overall architecture of our model. The model in this figure is generating target tokens of $M+1^\textit{th}$ bar. }
    \label{fig:model}
    \vspace{-0.5cm}
\end{figure}

\subsection{Modeling One Note in One Step}
\label{sec:note_level_modeling}

To shorten the token sequence to help the model learn from longer music structure, we apply note-level modeling to model multiple attributes of one note in one sequence step.
Different from previous works~\cite{huang2020pop,huang2018music,simon2017performance} which use multiple tokens to represent the attributes of one music note (pitch, velocity and duration), we regard each attribute of a note as an embedding and take the sum of all attribute embeddings as the representation of this note and take as input to the encoder and decoder in our sequence-to-sequence model in each timestep. When predicting multiple attributes of a music note, we add multiple softmax matrices on the output hidden of this step to generate corresponding attributes of this note. In this way, our input and output sequences can be much shorter, which can help our model better capture the long-term dependency. As a byproduct, it can also speed up the training and inference. We will describe the detailed implementation in Section \ref{input_output_module}.

\subsection{Modeling Long-Term Structure}

To capture the long-range context in sequence-to-sequence modeling, a lot of works~\cite{mikolov2012context,wang2015larger,zhang2018improving} directly feeds a representation of wider context into the model as an additional input in the encoder and decoder, but they çannot fully exploit the context~\cite{zheng2020toward} and are hard to maintain very long memory. Recently, \citet{dai2019transformer} propose a Transformer-based architecture called Transformer-XL which can learn dependency that is much longer than RNNs and vanilla Transformer and generate reasonably coherent and novel text articles with thousands of tokens. \citet{zheng2020toward} further extend Transformer-XL as the encoder and decoder model in document-level machine translation task and demonstrate the benefit of context far beyond the neighboring two or three sentences. Inspired by Transformer-XL, we enhance our encoder-decoder framework with recurrence Transformer encoder and recurrence Transformer decoder to model repetition and long-term structure in a musical piece. We regard one bar as one segment in our model.

\paragraph{Recurrent Transformer Encoder}

Recurrent Transformer encoder is very similar to Transformer-XL, which is used to encode each token $x_i$ in conditional tracks in one sequence step $i$. During training, the hidden state sequence computed in previous tokens is taken as memory and fixed, which is reused as an extended context by the recurrent Transformer encoder. The outputs of the encoder are then fed into the recurrent Transformer decoder as condition context. Recurrent Transformer encoder $\textit{Enc}$ and the outputs of the encoder $C_{i}$ can be formulated as follows: 
\begin{equation}
\begin{aligned}
C_{i} = \textit{Enc}(x_i, M^E_{i}),
\end{aligned}
\end{equation}
where $M^E_{i}$ represents the encoder memory used for $i$-th sequence step token, which is the encoder hidden state sequence computed in previous sequence steps.

\paragraph{Recurrent Transformer Decoder}

The decoder aims to generate token $y_j$ conditioned on the previously generated  tokens $y_{t<j}$ and context from encoder. During the training process, we apply bar-level attention mask to cross attention module to ensure that each token in the decoder can only see the condition context of the same bar. Recurrent Transformer decoder $\textit{Dec}$ and the decoder output $y_{j}$ can be formulated as follows:

\begin{equation}
\begin{aligned}
y_{j} = \textit{Dec}(y_{t<j}, M^D_{j}, C_{i \in Bar_j}),
\end{aligned}
\end{equation}
where $M^D_{j}$ represents the decoder memory used for $j$-th sequence step token which is the decoder hidden state sequence computed in previous sequence steps. $C_{i \in Bar_j}$ represents all encoder outputs in the bar where the generated token $y_j$ is located.

\subsection{Model Implementation}
In this subsection, we introduce the details of model implementation, including the input and output module, and the training and inference process.

\begin{figure*}
    \centering
    \includegraphics[width=0.95\textwidth, trim={0cm 10.5cm 2.2cm 0cm}, clip=true]{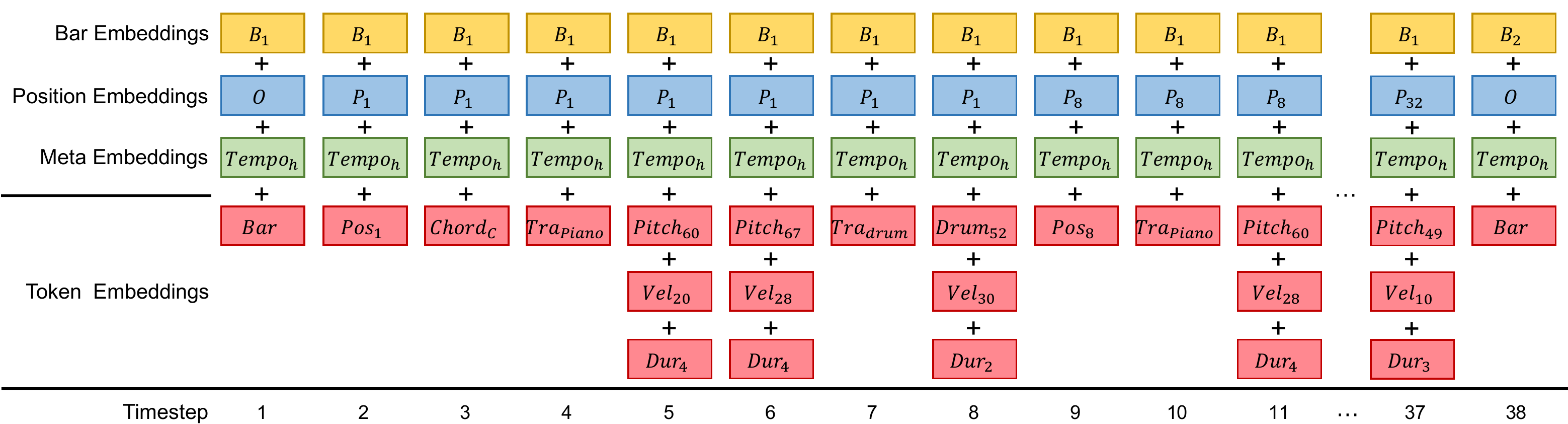}
    \vspace{-0.1cm}
    \caption{The input module of our model. The input embedding in each timestep is the sum of token embeddings, meta embedding, position embedding and bar embedding in this timestep. The input module converts the required MuMIDI symbols in the source/target side into input representations and then they are fed into the encoder/decoder. 
    }
    \label{fig:input_repr}
    \vspace{-0.4cm}
\end{figure*}

\begin{figure}
    \centering
    \includegraphics[width=0.42\textwidth, trim={0cm 0cm 0cm 0cm}, clip=true]{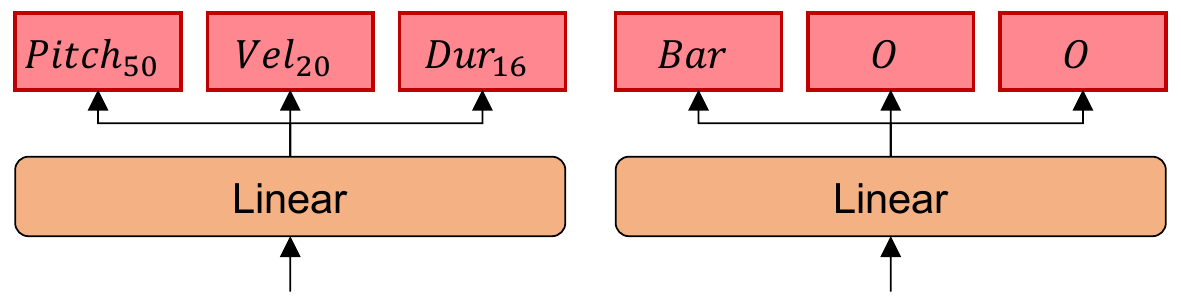}
    \vspace{-0.1cm}
    \caption{The output module in our model. The left subfigure shows an example when output module predicts a note symbol while the right subfigure shows when predicting a non-note symbol, such as bar, position and so on.}
    \label{fig:output_module}
    \vspace{-0.5cm}
\end{figure}

\subsubsection{Input and Output Module}
\label{input_output_module}

The input module is used to transform MuMIDI symbols into input representations. As shown in Figure \ref{fig:input_repr}, the input module consists of bar embeddings, position embeddings and token embeddings. We replace the positional encodings in vanilla Transformer and Transformer-XL with the combination of learnable bar embeddings and position embeddings to better make use of the order of the tokens. Compared to those commonly adopted positional encodings (such as those in vanilla Transformer and Transformer-XL), our bar and position embeddings 1) treat the notes in the same timestep equally, since these notes are performed simultaneously; and 2) make use of the order of the notes and distinguish the notes in different timesteps. We introduce each embedding in detail as follows.

\paragraph{Token Embeddings}
MuMIDI tokens contain the following types: note, bar, position, track, chord, etc. As mentioned in Section \ref{sec:note_level_modeling}, different from note token embeddings commonly used in previous works, we propose token embeddings that encode all attributes of one note (pitch, duration and velocity) into one token: we sum all embeddings of all attributes together as one sequence step. For other types (e.g., \textit{Bar}, \textit{Position}, \textit{Track}, \textit{Chord}) of MuMIDI tokens, we use one embedding to represent each of them.

\paragraph{Bar Embeddings}
Bar embeddings are used to indicate which bar the input token is located in. Bar embeddings are denoted as $B_1, ... B_m$, where $m$ is defined as the maximum number of bars in a music piece. When the number of bars exceeds $m$, we use $B_m$ as the bar embedding for those tokens in the bars out of $m$. 

\paragraph{Position Embeddings}
Position embeddings are used to indicate which timestep the current input token is located in. Position embeddings are denoted as $P_1, ... P_{32}$ and $O$, corresponding to 32 timesteps plus an empty symbol as mentioned in Section \ref{sec:bar_pos}. 

\paragraph{Meta Embeddings}
Meta embeddings encode meta symbols and then are added to tokens in all sequence step. In this paper, we only introduce tempo symbol as an example: we classify the tempo into three categories: low, middle and high as mentioned in Section \ref{sec:bar_pos}, corresponding to $\textit{Tempo}_{l}$, $\textit{Tempo}_{m}$ and $\textit{Tempo}_{h}$.

We then design a special output module to decode the attributes for notes, as shown in Figure \ref{fig:output_module}. Output module linearly projects the outputs of decoder $H_{dec}$ three times to obtain three different logits $H_{1}$, $H_{2}$ and $H_{3}$. We apply softmax function on them to yield three categorical probability distribution $D_{1}$, $D_{2}$ and $D_{3}$ over the note attributes (pitch, velocity and duration) of each note. If the symbol type of this sequence step is not \textit{note} (e.g., \textit{Bar}, \textit{Position}, \textit{Track}, \textit{Chord}), we only use $D_1$ as output and ignore $D_{2}$ and $D_{3}$ in training and inference.

\subsubsection{Training and Inference}

Finally, we describe the training and inference procedure of PopMAG according to the formulations in the previous subsections. 

In the training process, we adopt teacher forcing strategy and feed the ground truth tokens into the decoder to generate next tokens. We minimize the cross entropy between generated tokens and ground truth tokens to optimize the model. In the inference process, we generate the tokens in the target side one by one. We store the current bar/position embedding during inference and when a bar/position token is generated, we update the current bar/position embedding according to the generated token. We describe the training and inference procedure in detail in supplementary materials.

\section{Experimental Setup}

\subsection{Datasets}

We evaluate the performance of PopMAG on three music datasets: a pop music subset of Lakh MIDI dataset (denoted as \textit{LMD})~\cite{raffel2016learning}, a pop music subset of FreeMidi (denoted as \textit{FreeMidi}) and an internal Chinese Pop MIDI Dataset (denoted as \textit{CPMD}). For \textit{LMD}, we first get the meta information from LMD-matched subset following \citet{dong2018musegan}, including style tags (such as pop, classical, country and so on) of MIDIs matched to the Million Song Dataset~\cite{bertin2011million}. Then we filter the MIDIs to only obtain pop styles. For \textit{FreeMidi}, we crawl all MIDIs in pop genre in the FreeMidi website\footnote{https://freemidi.org/genre-pop}.

Since most MIDIs are user-generated and these datasets are too noisy for multi-track music accompaniment generation, we perform these cleansing and processing steps:
\begin{itemize}[leftmargin=*]
    \item \textbf{Melody Extraction}: Melody track is very important in pop music generation and is a fundamental part of pop music. However, melody track is usually played with different instruments and therefore we cannot simply choose a track as melody track according to its instrument or track name. To solve this issue, we use MIDI Miner~\cite{guo2019midi} to recognize the melody track. If a melody is not recognized by MIDI Miner, we choose the track performed by the flute as the melody track since the flute performs melody in most cases.
    \item \textbf{Track Compression}:  We compress other tracks into five types of tracks: bass, drum, guitar, piano and string, following ~\citet{dong2018musegan}. For bass track, if multiple bass tracks are overlapped, we choose the track with the most notes as the bass track.
    \item \textbf{Data Filtration}: First we filter tracks which contain less than 20 notes. After the track filtration, we then only keep MIDIs which a) contain at least 3 tracks; b) must contain melody track and at least one another track.
    \item \textbf{Data Segmentation}: We only consider 4/4 time signature in our implementation and thus we split each MIDI on each time change event and only keep those segments with 4/4 time signature which is the most commonly used time signature. 
    \item \textbf{Chord Recognition}: We infer chords with the Viterbi algorithm and use the implementation from Magenta\footnote{\url{https://github.com/tensorflow/magenta/blob/master/magenta/music/chord_inference.py}}. We infer two chords for each bar.
\end{itemize}

After cleansing and processing, we get 21916 musical pieces in \textit{LMD}, 5691 in \textit{FreeMidi} and 5344 in \textit{CPMD}. More detailed statistics of these datasets are shown in Table \ref{tab_dataset_stats}. Finally, we randomly split each dataset into 3 sets: 100 samples for validation, 100 samples for testing and the remaining samples for training.

\begin{table}[!h]
\centering
\vspace{-0.2cm}
\caption{The statistics of the datasets we used.}
\vspace{-0.3cm}
\begin{tabular}{ l | c | c  | c }
	\toprule
		Dataset &  \#Musical Pieces  & \#Bars & Duration (hours)  \\
		\midrule
		\textit{LMD} & 21916 & 372339  & 255.13  \\
		\textit{FreeMidi} & 5691 & 92825 & 52.32 \\
		\textit{CPMD} & 5344 & 94170 & 54.12  \\
		\bottomrule
\end{tabular}
\label{tab_dataset_stats}
\vspace{-0.2cm}
\end{table}

\subsection{Model Configurations}

Our model consists of a recurrent Transformer encoder and a recurrent Transformer decoder. We set the number of encoder layers, decoder layers, encoder heads and decoder heads to 4, 8, 8 and 8 respectively\footnote{We use a smaller encoder than the decoder because the target side has more tracks than the source side.}. The hidden size of all layers and the dimension of token, bar and position embeddings are set to 512. The length of training input tokens and the memory length are also to 512. The total number of learnable parameters is ~49M. We list detailed configurations in supplementary materials.

\subsection{Training and Evaluation Setup}
We train our model on 2 NVIDIA 2080Ti GPUs, with batch size of 10 musical pieces on each GPU. We use the Adam optimizer with $\beta_{1}= 0.9$, $\beta_{2} = 0.98$, $\varepsilon = 10^{-9}$ and follow the same learning rate schedule in \citep{vaswani2017attention}. It takes 160k steps for training until convergence. We regard generating five tracks (bass, piano, guitar, string and drum) conditioned on melody and chord (denoted as ``melody-to-others") as the default task unless otherwise stated. We set the maximum number of generated bars to 32. To ensure the diversity of generated musical pieces, we use stochastic sampling method during inference following most music generation systems~\cite{huang2020pop,huang2018music}.

\begin{figure*}[h] 
	\centering
	\vspace{-0.0cm}
	\begin{subfigure}[h]{0.28\textwidth}
		\centering
		\includegraphics[width=\textwidth, trim={-1cm 0cm -1cm 0cm}, clip=true]{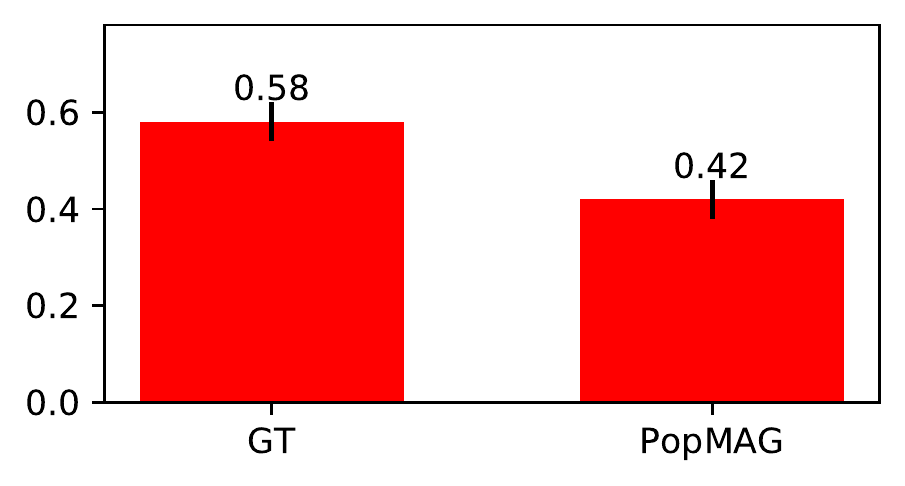}
		\caption{Preference scores on LMD. }
	\end{subfigure}
	\hspace{0.2cm}
	\begin{subfigure}[h]{0.28\textwidth}
		\centering
		\includegraphics[width=\textwidth, trim={-1cm 0cm -1cm 0cm}, clip=true]{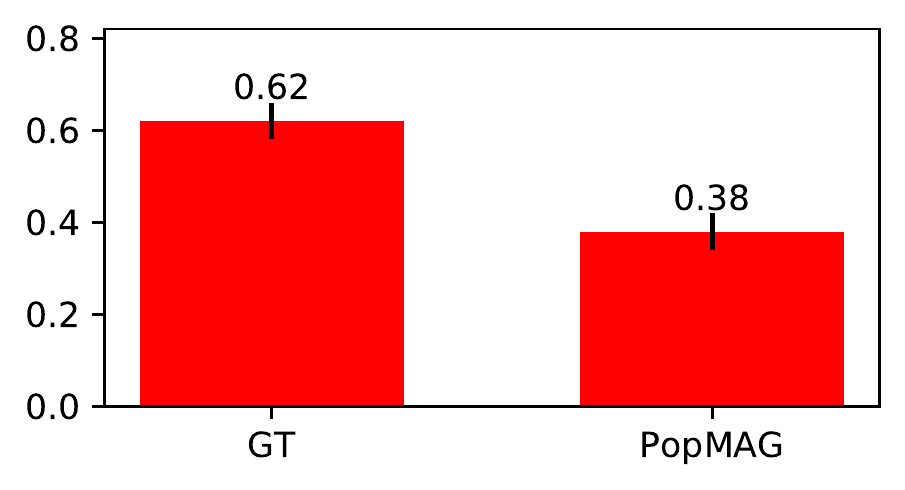}
		\caption{Preference scores on FreeMidi.}
	\end{subfigure}
	\hspace{0.2cm}
	\begin{subfigure}[h]{0.28\textwidth}
		\centering
		\includegraphics[width=\textwidth, trim={-1cm 0cm -1cm 0cm}, clip=true]{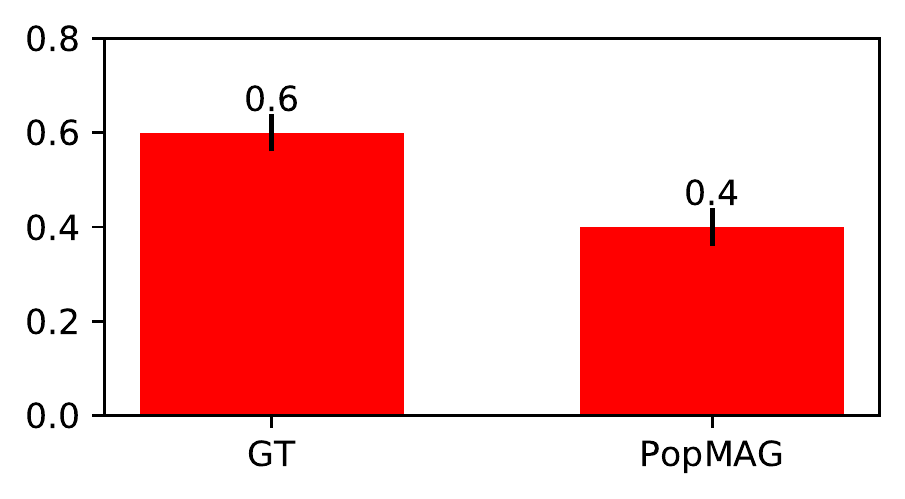}
		\caption{Preference scores on CPMD.}
	\end{subfigure}
	\vspace{-0.35cm}
	\caption{Subjective evaluations of PopMAG in melody-to-others task. Error bars show standard deviations of mean.}
	\label{fig:subjective_datasets}
	\vspace{-0.45cm}
\end{figure*}

\subsection{Evaluation Metrics}
We conduct both subjective and objective evaluations to measure the quality of the generated musical pieces. For objective evaluation, since the generation process is not deterministic due to stochastic sampling, we repeat each experiment 10 times on test set and calculate the average and 95\% confidence intervals for each objective metric\footnote{PPL is evaluated only once on the validation set with teacher-forcing strategy as that used in the training process.}.

\subsubsection{Subjective Evaluation}

Considering the diversity of generated music, there is not a suitable quantitative metric to evaluate the generation result. Thus, we validate the performance of methods based on human study. We ask totally 15 participants to evaluate the musical pieces. Among them, 5 evaluators can understand basic music theory. We pack the musical pieces from several settings (e.g., generated, ground truth) with the same conditional track together as one listening set and we have totally 100 listening sets, corresponding to 100 test musical pieces. Each listening set is evaluated by all participants and they are asked to choose the musical piece they like more by overall harmony. We average the total winning votes for each setting to obtain the final preference score.

\subsubsection{Objective Evaluation}
Objective evaluation in music generation remains an open question, though a variety of metrics have been proposed to evaluate the harmony, quality and similarity between one musical piece and another. Inspired by \citet{yang2018evaluation,zhu2018xiaoice,huang2020pop}, we propose the following metrics to evaluate the generated music.

\paragraph{Chord Accuracy (CA)} Chord Accuracy measures whether the chords of generated tracks match the conditional chord sequence, which affects the harmony of the generated music. Chord accuracy is defined as: 
\begin{equation*}
\begin{aligned}
    \textit{CA} & = \frac{1}{N_{\textit{tracks}}*N_{\textit{chords}}}\sum_{i=1}^{N_{\textit{tracks}}} \sum_{j=1}^{N_{\textit{chords}}}\mathbf{I}\{C_{i,j} = \hat{C}_{i,j}\},
\end{aligned}
\end{equation*}where $N_{\textit{tracks}}$ is the number of tracks, $N_{\textit{chords}}$ is the number of bars, $C_{i,j}$ denotes the $j$-th conditional (ground-truth) chord sequence in $i$-th track and $\hat{C}_{i,j}$ represents the generated $j$-th chord in $i$-th track.

\paragraph{Perplexity (PPL)}
Perplexity is a very common metric in text generation tasks~\cite{dai2019transformer,al2019character,baevski2018adaptive} to measure how good a model can fit the sequence, and it is also widely used to measure the performance of music generation~\cite{huang2018music,choi2019encoding}. We evaluate token-wise PPL on the validation set.

\paragraph{Pitch, Velocity, Duration and Inter-Onset Interval}
To further quantify the harmony, dynamics and expressiveness of a musical piece, we calculate the distribution of some features (e.g., pitch, velocity) and measure the distance of these distributions between the generated and ground-truth musical pieces. To get the distributions, we compute the histograms of each feature, then apply kernel density estimation~\cite{dehnad1987density} to convert the histograms into PDFs, which can smooth the histogram results for a more generalizable representation. The features we choose are listed as follows: 

\begin{itemize}[leftmargin=*]
    \item Pitch ($\textit{P}$): We compute the distribution of pitches classes.
    \item Velocity ($\textit{V}$): We quantize the note velocity into 32 classes corresponding to 32 velocity levels in note symbol and compute the distribution of classes.
    \item Duration ($\textit{D}$): To extract the note duration histogram, we quantize the duration into 32 classes corresponding to 32 duration attributes in note symbol and compute the distribution of classes.
    \item Inter-Onset Interval ($\textit{IOI}$): Inter-onset interval~\citet{yang2018evaluation} is the interval between two note onsets. We quantize the intervals into 32 classes the same as note duration and compute the distribution of interval classes.
\end{itemize}

We then compute the averaging Overlapped Area ($\textit{OA}$) of distributions ($\mathcal{D}_\mathcal{A}$, $\mathcal{A}$ can be one of $\textit{P}$, $\textit{V}$, $\textit{D}$ and $\textit{IOI}$) to measure the difference between generated musical piece and ground-truth musical piece:

\begin{equation*}
    \mathcal{D}_\mathcal{A} = \frac{1}{N_{\textit{tracks}}*N_{\textit{bars}}}\sum_{i=1}^{N_{\textit{tracks}}} \sum_{j=1}^{N_{\textit{bars}}}\textit{OA}(\mathcal{P}^\mathcal{A}_{i,j}, \hat{\mathcal{P}}^\mathcal{A}_{i,j})
\end{equation*} 
where \textit{OA} represents the averaging overlapped area of two distributions, $\mathcal{P}^{\mathcal{A}}_{i,j}$ denotes the distribution of feature $\mathcal{A}$ in $i$-th bar and $j$-th track in ground truth musical piece, and $\hat{\mathcal{P}}^{\mathcal{A}}_{i,j}$ denotes that in generated musical piece.

\section{Results and Analyses}

\subsection{Overall Quality} To evaluate the overall harmony and high-quality of musical pieces generated by PopMAG, we first conduct subjective evaluations on LMD, FreeMidi and CPMD datasets. The results are shown in Figure \ref{fig:subjective_datasets}. We can see that although there is still a gap between ground-truth (human-composed) and generated musical pieces, a large part of generated musical pieces (about 42\%, 38\%, 40\% in three datasets) have reached the quality of ground-truth ones, which demonstrates that PopMAG is quite promising to generate expressive and contagious accompaniments.

\subsection{Comparison with Previous Works}
\label{sec:compare_with_musegan}

We compare the music quality of our model with another multi-track accompaniment generation model: MuseGAN~\cite{dong2018musegan}\footnote{We do not compare our model with XiaoIce Band~\cite{zhu2018xiaoice} because its codes were not released yet and complicated to reproduce.}. To make our model and MuseGAN comparable 1) we conduct experiments in the same task (the task used in MuseGAN, generating four tracks (guitar, drum, string and bass) conditioned on piano track) and use same training/test splits for both models; 2) both models are asked to generate 4 bars of notes in target tracks, since MuseGAN cannot generate longer musical pieces; 3) both models do not use chord since MuseGAN does not introduce chord condition and 4) we set the velocity of all notes in musical pieces generated by PopMAG and MuseGAN to a reasonable value (100). The results are shown in Table \ref{tab:results_other_works} and Figure \ref{fig:results_other_works}. We can see that PopMAG outperforms MuseGAN on all subjective and objective metrics, demonstrating the high quality of musical pieces generated by PopMAG and that our sequence representation (MuMIDI) can keep harmony better than pianoroll-based representation (MuseGAN). Besides, PopMAG can generate very long musical pieces while MuseGAN can only generate short and fix-length musical pieces. 

\begin{table}[!htb]
\small
\centering
\vspace{-0.2cm}
\caption{The comparison of performances between PopMAG and MuseGAN on LMD dataset in piano-to-others task. The best number of each metric is marked in bold.}
\vspace{-0.2cm}
\begin{tabular}{ l | c | c | c | c  }
	\toprule
	 & \textit{CA} & $\mathcal{D}_\textit{P}$ & $\mathcal{D}_\textit{D}$  & $\mathcal{D}_\textit{IOI}$  \\
	\midrule
	\textit{MuseGAN}~\cite{dong2018musegan}  
	& 0.37 $\pm$ 0.02 & 0.21 $\pm$ 0.01 & 0.35 $\pm$ 0.01 & 0.28 $\pm$ 0.02 \\
  	\textit{PopMAG}  
  	& \textbf{0.45 $\pm$ 0.01} & \textbf{0.58 $\pm$ 0.01} & \textbf{0.55 $\pm$ 0.01} & \textbf{0.72 $\pm$ 0.01} \\
	\bottomrule
\end{tabular}
\label{tab:results_other_works}
\vspace{-0.5cm}
\end{table}

\subsection{Method Analyses}
\subsubsection{Comparison with Other MIDI Representation}
\begin{figure*}[h] 
	\centering
	\begin{subfigure}[h]{0.23\textwidth}
		\centering
		\includegraphics[width=\textwidth, trim={0cm 0cm 0cm 0cm}, clip=true]{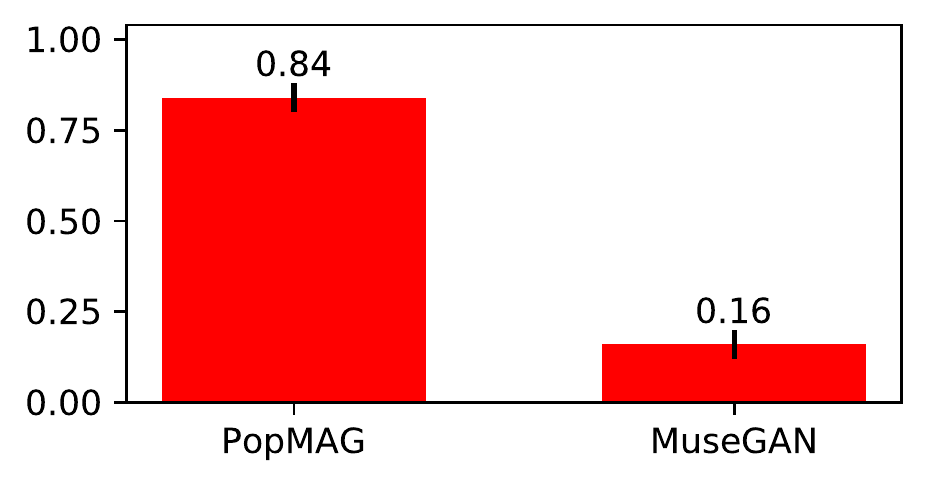}
		\vspace{-0.5cm}
		\caption{Preference scores of MuseGAN and PopMAG. }
		\label{fig:results_other_works}
	\end{subfigure}
	\hspace{0.2cm}
	\begin{subfigure}[h]{0.23\textwidth}
		\centering
		\includegraphics[width=\textwidth, trim={0cm 0cm 0cm 0cm}, clip=true]{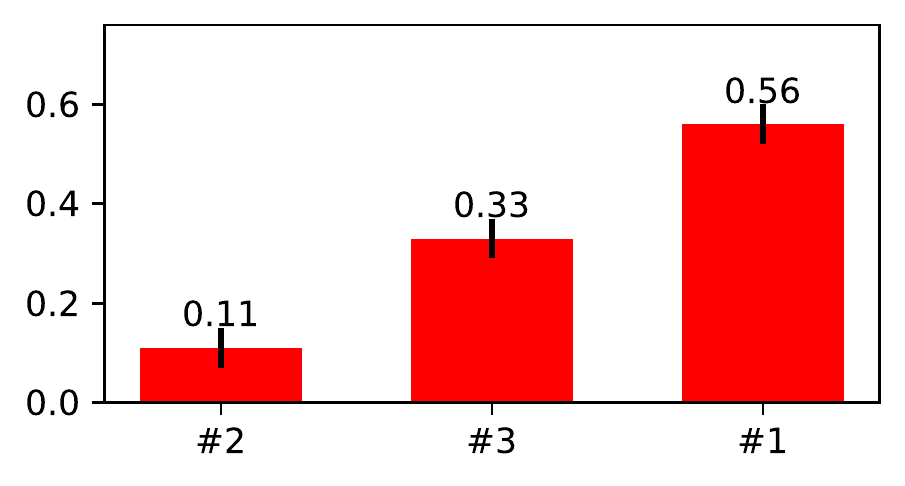}
		\vspace{-0.5cm}
		\caption{Preference scores of different MIDI representations.}
		\label{fig:subj_repre}
	\end{subfigure}
	\hspace{0.2cm}
	\begin{subfigure}[h]{0.23\textwidth}
		\centering
		\includegraphics[width=\textwidth, trim={0cm 0cm 0cm 0cm}, clip=true]{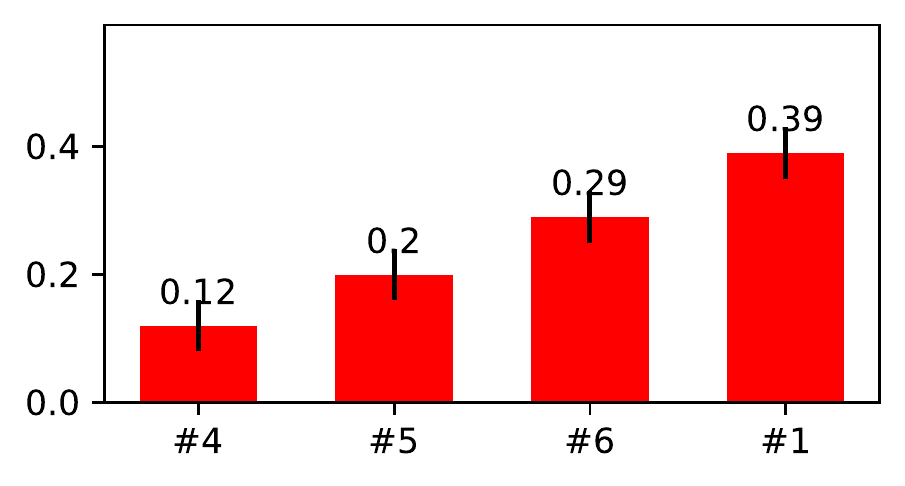}
		\vspace{-0.5cm}
		\caption{Preference scores of different context memory.}
		\label{fig:memory}
	\end{subfigure}	
	\hspace{0.2cm}
	\begin{subfigure}[h]{0.23\textwidth}
		\centering
		\includegraphics[width=\textwidth, trim={0cm 0cm 0cm 0cm}, clip=true]{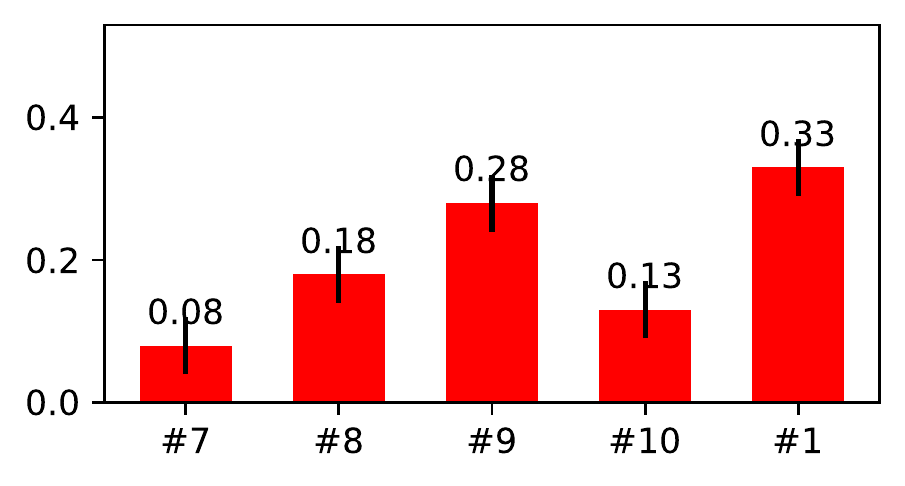}
		\vspace{-0.5cm}
		\caption{Preference scores of different position embeddings.}
		\label{fig:position_embed}
	\end{subfigure}
	\vspace{-0.3cm}
	\caption{Subjective evaluations of several settings. Error bars show standard deviations of mean. All settings are evaluated on LMD in melody-to-others task except (a) which is evaluated in piano-to-others task.}
	\label{fig:subjective}
	\vspace{-0.3cm}
\end{figure*}

To analyze the effectiveness of MuMIDI representation, we compare MuMIDI with other commonly used MIDI representations including MIDI-like representation~\cite{huang2018music} and REMI~\cite{huang2020pop}. To make these models support multi-track music accompaniment generation, we do some modification on them: for MIDI-like representation, we extend it to multi-track version following LakhNES~\cite{donahue2019lakhnes} which use different tokens to represent notes of different instruments with the same pitch and also add chord symbols to the conditional sequence  (denoted as \textit{MIDI-like}); for REMI, we add a track token before each \textit{Note On} token to represent the track of this note (denoted as \textit{REMI}). For all representations, we use the same accompaniment generation model with context memory in the encoder and decoder. We conduct subjective and objective evaluation on three systems and the results are shown in Table \ref{tab:results_all} (from Row 1 to Row 3) and Figure \ref{fig:subj_repre}. We can see that \textit{MuMIDI} achieves better scores than \textit{MIDI-like} and \textit{REMI} on both subjective and objective metrics, indicating that MuMIDI can generate more harmonious musical piece than other MIDI representations.

\subsubsection{Analyses on Note-Level Modeling}
To verify the effectiveness of our note-level modeling method (modeling one note in one step), we report the average length of target token sequences in the training set, training time and inference latency in Table \ref{tab:results_repr_latency}. The results show that our model can converge faster and generate musical piece faster than other MIDI representation modeling methods, thanks to the shorter token sequence with note-level modeling.

\begin{table}[!h]
\centering
\vspace{-0.2cm}
\caption{The comparison of the average length of target token sequence, training time and inference latency among PopMAG and other MIDI representation modeling methods on LMD dataset in melody-to-others task.}
\vspace{-0.2cm}
\label{tab:results_repr_latency}
\begin{tabular}{ l | c | c | c }
	\toprule
	  Settings & Average Length & \tabincell{c}{Training Time \\ (hours)} & \tabincell{c}{Latency\\ (s/bar)} \\
	\midrule
	\textit{MIDI-like}~\cite{huang2018music,donahue2019lakhnes}  & 3108 & 85 & 1.59 \\
   	\textit{REMI}~\cite{huang2020pop}  & 3478 & 81 & 1.57 \\ \midrule
   	\textit{MuMIDI}  & \textbf{1805} & \textbf{56} & \textbf{0.75} \\
	\bottomrule
\end{tabular}
\vspace{-0.4cm}
\end{table}

\begin{table*}[!htb]
\centering
\small
\caption{The results comparison of among different settings of PopMAG on LMD dataset in melody-to-others task. We use the same model for all MIDI representations for a fair comparison. Row 2 and 3 show the performance of other MIDI representations. Row 4 to 6 study the context memory in our encoder and decoder. Row 7 to 10 explore different position embeddings in our input module. The best number of each metric is marked in bold.}
\vspace{-0.2cm}
\begin{tabular}{c | l | c | c | c | c | c | c  }
	\toprule
	No. & Settings & \textit{CA} & \textit{PPL} & 
	$\mathcal{D}_\textit{P}$ & $\mathcal{D}_\textit{V}$ & 
	$\mathcal{D}_\textit{D}$  & $\mathcal{D}_\textit{IOI}$   \\
	\midrule
	\#1 & \textit{PopMAG}  
	& \textbf{0.647 $\pm$ 0.013} & \textbf{1.131} & \textbf{0.602 $\pm$ 0.012} & \textbf{0.454 $\pm$ 0.007} & \textbf{0.478 $\pm$ 0.010} & \textbf{0.688 $\pm$ 0.007} \\ \midrule \midrule
   	\#2 & \textit{REMI}~\cite{huang2020pop}  
   	& 0.552 $\pm$ 0.019 & /     & 0.406 $\pm$ 0.009 & 0.345 $\pm$ 0.147 & 0.387 $\pm$ 0.011 & 0.557 $\pm$ 0.011   \\
   	\#3 & \textit{MIDI-Like} 
   	& 0.181 $\pm$ 0.020 & /     & 0.307 $\pm$ 0.019 & 0.305 $\pm$ 0.147 & 0.364 $\pm$ 0.013 & 0.498 $\pm$ 0.012  \\ \midrule
	\#4 & \textit{PopMAG} - \textit{DM} - \textit{EM} 
	& 0.617 $\pm$ 0.024 & 1.143 & 0.531 $\pm$ 0.006 & 0.418 $\pm$ 0.008 & 0.453 $\pm$ 0.009 & 0.661 $\pm$ 0.006 \\
    \#5 & \textit{PopMAG} - \textit{DM} 
    & 0.634 $\pm$ 0.009 & 1.139 & 0.581 $\pm$ 0.002 & 0.453 $\pm$ 0.009 & 0.476 $\pm$ 0.009 & 0.683 $\pm$ 0.009 \\ 
    \#6 & \textit{PopMAG} - \textit{EM}  
    & 0.642 $\pm$ 0.016 & 1.135 & 0.582 $\pm$ 0.012 & \textbf{0.454 $\pm$ 0.012} & \textbf{0.478 $\pm$ 0.008} & 0.681 $\pm$ 0.007  \\ \midrule
	\#7 & \textit{PopMAG} - \textit{POS} - \textit{BAR}
	& 0.483 $\pm$ 0.024 & 1.272 & 0.350 $\pm$ 0.012 & 0.201 $\pm$ 0.006 & 0.302 $\pm$ 0.010 & 0.520 $\pm$ 0.009  \\
   	\#8 & \#7 +\textit{Sinusoidal}~\cite{vaswani2017attention}  
   	& 0.636 $\pm$ 0.014 & 1.163 & 0.563 $\pm$ 0.006 & 0.435 $\pm$ 0.008 & 0.463 $\pm$ 0.006 & 0.671 $\pm$ 0.003  \\
   	\#9 & \#7 +\textit{Relative}~\cite{dai2019transformer}  
   	& 0.641 $\pm$ 0.009 & 1.144 & 0.582 $\pm$ 0.005 & 0.439 $\pm$ 0.009 & 0.469 $\pm$ 0.005 & 0.678 $\pm$ 0.005  \\
    \#10 & \#7 +\textit{POS}   
    & 0.610 $\pm$ 0.012 & 1.152 & 0.530 $\pm$ 0.007 & 0.385 $\pm$ 0.012 & 0.431 $\pm$ 0.008 & 0.667 $\pm$ 0.008  \\
	\bottomrule
\end{tabular}
\label{tab:results_all}
\vspace{-0.2cm}
\end{table*}

\subsubsection{Analyses on Memory in the Encoder and Decoder}

To investigate the effectiveness of the context memory in the encoder and decoder, we compare \textit{PopMAG} with three settings:
1) \textit{PopMAG - DM - EM}, which removes memory in the encoder and decoder of PopMAG; 
2) \textit{PopMAG - DM}, which only removes memory in the decoder; 
3) \textit{PopMAG - EM}, which only removes memory in the encoder.
The results are shown in Table \ref{tab:results_all} (Row 1 and Row 4 to Row 6) and Figure \ref{fig:memory}, we can see that 1) \textit{PopMAG} performs best in all metrics, demonstrating that memory in the encoder and decoder improves the model performance. 2) \textit{PopMAG - EM} outperforms \textit{PopMAG - DM} in most metrics, indicating that context memory in the decoder is more important than that in the encoder; and 3) both memory in the encoder and decoder can help long-term modeling and improve the harmony of generated musical pieces.

\subsubsection{Analyses on Bar and Position Embeddings}
To prove the effectiveness of our bar and position embeddings, we compare them with sinusoidal~\cite{vaswani2017attention} and relative~\cite{dai2019transformer} position encodings. We list the results in Row 7 to Row 10 in Table \ref{tab:results_all} and Figure \ref{fig:position_embed}. Compare them with Row 1 which uses our bar and position embeddings, we can see that: 1) The combination of bar and position (\#1) embeddings outperforms sinusoidal (\#8) and relative position encodings (\#9), which demonstrates that our bar and position embeddings can help model capture the music structure better. 2) Our bar embeddings (\#10 and \#1) and position embeddings (\#7 and \#10) are both helpful for modeling the position. The results of the subjective evaluation are also consistent with the above analysis.

\subsection{Extension}
PopMAG is general and can be applied in different settings for music accompaniments generation: 1) generating from scratch, which generates multi-track accompaniments conditioned only on melody and chord; 2) starting from semi-manufactured music, which enriches a song with more expressive accompaniment tracks by generating more tracks conditioned on melody, chord, and a few tracks; and 3) recomposing a song, which removes some tracks and generates some other tracks. Furthermore, PopMAG can be combined with singing voice synthesis system~\cite{ren2020deepsinger,lee2019adversarially,lu2020xiaoicesing} to produce a whole pop song.

\section{Conclusion}
In this work, we proposed PopMAG, a pop music accompaniment generation framework, to address the challenges of multi-track harmony modeling and long-term dependency modeling in music generation. PopMAG includes a novel MUlti-track MIDI representation (MuMIDI) which encodes multi-track MIDI events into a single sequence and an enhanced sequence-to-sequence model with note-level modeling and extra-long context. Experiments on multiple datasets (LMD, FreeMidi and an internal dataset of Chinese pop songs) demonstrate the effectiveness of PoPMAG for multi-track harmony and long-term dependency modeling. 

In the future, we will study fine-granularity music accompaniment generation and integrate emotion and style-controlled generations into PopMAG. We will also consider large-scale generative pre-training~\cite{brown2020language,song2019mass} to improve the generation quality. Furthermore, we will apply PopMAG to other tasks such as chord progression generation, singing voice accompaniment generation and MIDI classification. PopMAG can be leveraged to improve the productivity of musicians, and inspire them to compose more high-quality accompaniments.

\section*{Acknowledgments}
This work was supported in part by the National Key R\&D Program of China (Grant No.2018AAA0100603), Zhejiang Natural Science Foundation (LR19F020006), National Natural Science Foundation of China (Grant No.61836002, No.U1611461 and No.61751209) and the Fundamental Research Funds for the Central Universities (2020QNA5024).

\newpage

\bibliographystyle{ACM-Reference-Format}
\bibliography{reference}

\newpage
\appendix
\section{Appendix}

\subsection{Model Settings} 
\label{sec:model_config}
We list the model settings of our model in Table \ref{tab:hp_seq2seqxl}.

\begin{table}[h]
\centering
\begin{tabular}{l|l}
\hline
\textbf{Model Setting}            &   \textbf{Value} \\ \hline
Token Embedding Dimension         &     512   \\ \hline
Encoder Layers                    &     4     \\ \hline
Decoder Layers                    &     8     \\ \hline
Encoder/Decoder Hidden            &     512   \\ \hline
Encoder/Decoder Filter Size       &    2048  \\ \hline
Encoder/Decoder Attention Heads   &     8     \\ \hline
Dropout                           &     0.1   \\ \hline \hline
Total Number of Parameters        &     49.01M  \\ \hline
\end{tabular}
\caption{Model settings of our model. }
\label{tab:hp_seq2seqxl}
\end{table}

\begin{algorithm}[!h]
\small
\caption{PopMAG Training}
\label{alg_popmag_train}
\begin{algorithmic}
    \State {\bfseries Input}: Multi-tracks musical pieces $(\mathcal{X}, \mathcal{Y})$ in MuMIDI representation where $\mathcal{X}$ represents conditional tracks and $\mathcal{Y}$ represents target tracks. 
    \State \textbf{Initialize}: Recurrent Transformer encoder $\textit{Enc}$, recurrent Transformer decoder $\textit{Dec}$, input module $I$, output module $P$, encoder memory $M^E$ with $\emptyset$, decoder hidden memory $M^D$ with $\emptyset$, maximum length of encoder memory $m_E$, maximum length of decoder memory $m_D$, and total training epoch $e$.
    \For{each $epoch\in [0,e)$}
        \For{each $(x^{\textit{MuMIDI}}, y^{\textit{MuMIDI}}) \in (\mathcal{X}, \mathcal{Y}$)}
            \State $x = I(x^{\textit{MuMIDI}})$, $y = I(y^{\textit{MuMIDI}})$
            \For{each $i\in [0,\#\textit{Bars}(y))$}
                \For{each $j \in [\textit{Bs}_i, \textit{Be}_i]$}
                    \State $C_{j}, M^E_{j} = \textit{Enc}(x_j, x_{\textit{Bs}_i \le t \le \textit{Be}_i}, \textit{SG}(M^E_{max(\textit{Bs}_i-m_E, 0) \le t \le \textit{Bs}_i}))$
                \EndFor
                \For{each $j \in [\textit{Bs}_i, \textit{Be}_i]$}
                    \State $y_{j}, M^D_{j} = \textit{Dec}(y_{t<j}, C_{\textit{Bs}_i \le t \le \textit{Be}_i}, \textit{SG}(M^D_{max(i-m_D, 0) \le t < j}))$
                    \State $y_i^{\textit{dist}} = P(O_{i})$
                    \State $ \textit{loss} = \textit{CrossEntropy}(y_i^{\textit{dist}}, y_i) $
                    \State Optimize $\textit{Enc}$, $\textit{Dec}$, $I$ and $P$ with \textit{loss}
                \EndFor
            \EndFor 
        \EndFor
    \EndFor
\end{algorithmic}
\end{algorithm}

\subsection{PopMAG Training and Inference}
\label{sec:training_and_infer}
During inference, we judge the first output sampled from categorical probability distribution $D_1$ in each sequence step. If it represents the pitch of a note, then we continue to take the second and the third tokens as velocity and duration and combine them as a note symbol. Otherwise, we only use the first token as a non-note symbol.
The detailed training and inference procedure is shown in Algorithm~\ref{alg_popmag_train} and \ref{alg_popmag_infer}.

\begin{algorithm}[!h]
\small
\caption{PopMAG Inference}
\label{alg_popmag_infer}
\begin{algorithmic}
    \State {\bfseries Input}: Conditional tracks of one multi-tracks musical piece $x^{\textit{MuMIDI}}$ in MuMIDI representation.
    \State \textbf{Load Model}: $\textit{Enc}$, $\textit{Dec}$, $I$, $P$.
    \State \textbf{Initialize}: Encoder memory $M^E$ with $\emptyset$, decoder hidden memory $M^D$ with $\emptyset$, maximum length of encoder memory $m_E$, maximum length of decoder memory $m_D$, number of generated bars $n=0$, number of maximum generated bars $max_b$.
    \For{each $i\in [0,\#\textit{Bars}(x))$}
        \For{each $j \in [\textit{Bs}_i, \textit{Be}_i]$}
        \State $C_{j}, M^E_{j} = \textit{Enc}(x_j, x_{\textit{Bs}_i \le t \le \textit{Be}_i},M^E_{max(\textit{Bs}_i-m_E, 0) \le t \le \textit{Bs}_i})$
        \EndFor
    \EndFor
    \State Set $y^{\textit{MuMIDI}}_0$ to the bar symbol. Set $y_0 = I(y^{\textit{MuMIDI}}_0)$. Set $j=1$.
    \While{$n < max_b$}
        \State $O_{j}, M^D_{j} = \textit{Dec}(y_{t<j}, C_{\textit{Bs}_n \le t \le \textit{Be}_n},  M^D_{max(j-m_D, 0) \le t < j})$
        \State $y^{\textit{dist}}_j = P(O_{j})$
        \State Sample $y_j$ from $y^{\textit{dist}}_j$ with top-k temperature-controlled stochastic sampling method~\cite{keskar2019ctrl}.
        \If{$y_i$ is a bar token}
            \State $n = n+1$. Update the current bar number in $I$ to $n$.
        \EndIf
        \If{$y_i$ is a position token}
            \State Update the current position number in $I$ to the position number indicated by $y_j$.
        \EndIf
        \State $j = j + 1$
    \EndWhile
\end{algorithmic}
\end{algorithm}

\end{document}